\documentstyle[aps,epsf,prb,multicol]{revtex}
\begin{document}
\newcommand{\Dvec}{{\bf D}}
\newcommand{\Evec}{{\bf E}}
\newcommand{\Pvec}{{\bf P}}
\newcommand{\alp}{\alpha}
\newcommand{\epsi}{\epsilon}
\newcommand{\evac}{\varepsilon_{\rm v}}
\newcommand{\ez}{\varepsilon^0}
\newcommand{\einf}{\varepsilon^\infty}
\newcommand{\sia}{$\partial\sigma/\partial a$}
\newcommand{\sic}{$\partial\sigma/\partial c$}
\newcommand{\siu}{$\partial\sigma/\partial u$}
\newcommand{\kvec}{{\bf k}}
\newcommand{\vv}{\!\!\!\!\!\!}
\newcommand{\vare}{\varepsilon}  
\newcommand{\eps}{\epsilon}  
\newcommand{\De}{$\Delta$}
\newcommand{\de}{$\delta$}
\newcommand{\mcol}{\multicolumn}
\newcommand{\be}{\begin{eqnarray}}
\newcommand{\ee}{\end{eqnarray}}
\draft
\title{Electronic dielectric constants of insulators 
  by the polarization method}

\author{Fabio Bernardini and  Vincenzo Fiorentini}

\address{Istituto Nazionale per la Fisica della Materia --
Dipartimento di Fisica, Universit\`a di Cagliari, Italy}

\date{June 3, 1998}
\maketitle

\begin{abstract}
We discuss a non-perturbative, technically straightforward,
 easy-to-use, and computationally affordable method, based 
on  polarization theory, for the
calculation of the electronic dielectric  constant of insulating
solids at the first principles 
level.  We apply the method to GaAs, AlAs, InN, SiC, ZnO, GaN,
AlN, BeO, LiF, PbTiO$_3$,  and CaTiO$_3$. 
The predicted  $\einf$'s  agree
well with  those given  by Density Functional Perturbation Theory
(the  reference theoretical treatment), and they are  generally
within less than   10 \% of experiment.
\end{abstract}


\begin{multicols}{2}
The electronic dielectric constant $\einf$, measuring the response to 
a uniform electrostatic field,  is a fundamental quantity in
basic and applied solid state physics. Besides
its intrinsic  interest, knowledge of $\einf$ is crucial to the
calculation of the static dielectric constant $\ez$. 
Ab initio calculations of $\einf$ have been  performed in recent times
by density functional perturbation theory (DFPT)
\cite{dfpt,dfpt2,rabe} for a host of different materials.
Here we discuss a new method to compute $\einf$
based on density functional theory \cite{dft} and the
modern theory of dielectric polarization,\cite{polariz} and
apply it to a set of polar crystals. Despite its simplicity and
ease of use,
the  method  predicts $\einf$'s in close agreement  with those
obtained by DFPT. This suggests that theoretical
predictions deviate from 
 (and typically overestimate)  the   experimental value  on 
account of density functional theory, and not of the specific 
method or implementation. 

We have sketched the basics of
the method in a report on the calculation of the
static dielectric constant.\cite{noi.prl}
Almost all  the ingredients needed to evaluate the latter
can be obtained from  ab-initio calculations of total energy, stress,
forces, and  dielectric polarization in zero field (by the Berry-phase
approach \cite{polariz,nota}) for bulk systems. The notable  exception
is the electronic dielectric constant $\varepsilon^\infty$,  for
which we follow the alternate approach described below. 

\paragraph*{Theory -- } 
The core of the argument is that  $\einf$ can be obtained from  the
relationship  
between macroscopic polarization in zero field and
interface  charge accumulation
\cite{noi.prl,noi.mrs,Vanderbilt.PRB48}
in  appropriate superlattices. 
An
insulating superlattice is constructed,
which consists of periodically alternating   
slabs of equal length, stacked along some fixed 
direction and made of materials 1 and 2.
In such a superlattice, in the absence of external sources of fields,
the displacement field orthogonal to the interfaces is conserved:
\begin{equation}
D_1 = E_1 + 4 \pi P_1 (E_1) =
 E_2 + 4 \pi P_2 (E_2) = D_2.\label{uno}
\end{equation}
We expand  the polarization to first order in the screened fields in
the two materials, indexed by $i$:
\begin{equation}
P_i (E)= P_i^{(0)} + \chi_i E_i, \label{due}
\end{equation}
where  $P_i^0$ is the {\it zero
 field} polarization and $\chi_i$ the susceptibility of material $i$.
The presence of a zero field polarization
is important: substituting the last relation into Eq. \ref{uno} one
 obtains 
\begin{equation}
 4 \pi (P_2^{(0)} - P_1^{(0)}) = \varepsilon^{\infty}_1 E_1 -
\varepsilon^{\infty}_2 E_2,\label{tre}
\end{equation}
(In this context, it does not matter whether the  zero field
 polarization is spontaneous or piezoelectric, or a combination of the 
two.) Only for a null $P_i^0$ does one recover the familiar equality  
\begin{equation}
\varepsilon^{\infty}_1 E_1 =
\varepsilon^{\infty}_2 E_2\,.
\end{equation}
To proceed further, one notes that 
periodic boundary conditions imply  $$E \equiv E_1=-E_2, \ \ \ \ \ \
\Delta E \equiv E_1 - E_2 = 2 E,$$ and therefore 
Eq. \ref{tre} becomes
\begin{equation}
4 \pi (P_2^{(0)} - P_1^{(0)}) = \frac{1}{2}(\varepsilon^{\infty}_1 +
\varepsilon^{\infty}_2) \Delta E.
\end{equation}
Recalling that the charge accumulation per unit  area
at the interface between materials 1 and 2 is
$s_{\rm int} = \pm \Delta E / 4\pi$, we finally obtain
\be \label{eq.einf}
  s_{\rm int}& = &\pm 2\, (P_2^{(0)} - P_1^{(0)}) /
(\varepsilon^{\infty}_1 + \varepsilon^{\infty}_2 ) \nonumber\\
 & =&
 \pm 2\, \Delta P^{(0)} /
(\varepsilon^{\infty}_1 + \varepsilon^{\infty}_2 ).
\ee
This relation  connects the {\it difference} in
 macroscopic bulk polarization at zero field
with  the components
$\varepsilon^{\infty}_{1,2}$
 of the dielectric tensors of the interfaced materials
along the interface  normal.\cite{nota2}

In an undistorted
 homojunction, i.e. a superlattice in which material 1 is identical to 
material 2, there is effectively no interface. Therefore
there is no polarization change, and  the  interface charge
is zero. It is nevertheless possible to generate a
 polarization difference  in a controlled manner,
 by  inducing  a small distortion $\delta$
of one of the atomic sublattices in half of the superlattice
unit cell. This produces a difference in polarization, and a charge
 accumulation at the interface.  The interface  charge $s_{\rm int}$
at the interface  between distorted and undistorted
regions can   be easily  calculated via macroscopic averaging
 \cite{noi.mrs,resta} of the charge density. On the other hand,
 the  zero field  polarizations $\Pvec_2$ for the bulk  material
in the undistorted state,
and $\Pvec_1$ for the bulk material in the same strain state
as in the superlattice, are evaluated using
the Berry phase technique. From Eq. (\ref{eq.einf}),
one then  extracts the
 average electronic dielectric constant $\bar\vare^{\infty} = 
(\varepsilon^{\infty}_1 + \varepsilon^{\infty}_2 )$/2.  

By construction $\vare_2$ equals the 
 dielectric constant, while  $\vare_1$, 
 the dielectric constant in the 
distorted state, of course does not. Therefore 
$\bar\vare$ does not equal the sought-after dielectric
constant for $\delta \neq 0$; However, 
$\bar\vare$  does equals the dielectric constant (the tensor component
 along $\hat{\bf n}$)
in the limit of zero distortion: $$
\varepsilon^{\infty} = \lim_{\delta\rightarrow 0}\, \bar\vare.$$
This limit can be evaluated  with essentially
arbitrary accuracy by extrapolation or interpolation.
Also, since the materials involved in
the heterojunction are identical modulo a
vanishingly small distortion, it is virtually guaranteed that no
interface  state exists, so that no band bending or electrostatic
perturbation  adulterates the polarization effects.

Summarizing, in the present approach the dielectric constant 
is simply obtained using the 
geometric quantum phase polarization and 
 relatively small, accurately controllable supercell
calculations. In the latter calculations, the slabs should  be short  
enough that the constant electric field will not cause metallization,
and the slabs should be sufficiently long to recover
bulk-like behavior away from the interfaces. Both requirements are
generally met also by materials with small calculated gaps 
if sufficiently small strains are applied. Unlike DFPT,
 the present method  does not resort to perturbation
theory, and it is novel in that  the determination
of the  electronic screening uses the connection 
 with the geometric quantum phase.\cite{resta2} 
The calculations involved are non-intensive, and can be performed even
on Pentium-like personal computers. Also, since the implementation is 
very much simpler than that of DFPT, the method seems  promising
as a general-purpose tool for non-specialists.

\paragraph*{Applications -- }
We now apply the method to a set of   representative polar
materials of general
interest: SiC, GaAs, AlAs, InN, GaN, AlN, BeO, ZnO, LiF,  cubic 
PbTiO$_3$,
and cubic CaTiO$_3$.
All our calculations, as detailed below, are done in the  local density 
approximation to density functional theory.\cite{dft} 
 Our results are 
compared to results of DFPT calculations (where available), which
effectively function as reference for new computational methods in 
this area. Comparison with experiment is also presented when possible.

In the calculations we use the Ceperley-Alder
exchange-correlation energy \cite{ca} and ultrasoft 
pseudopotentials \cite{USPP} for the electron-ion interaction.
The pseudopotentials have been constructed to include the
following semicore states in the valence manifold:  Zn and Ga 3$d$, 
In 4$d$; Li 1$s$; Pb 5$d$, Ti and Ca 3$s$3$p$.
 A  plane-wave basis
with 25 Ryd cutoff is found to be sufficient to converge the
quantities of interest in all the materials investigated.
Bulk Brillouin zone summations are done over appropriate
Chadi-Cohen \cite{CC} k-point meshes for the relevant structures.
Bulk polarizations are obtained  in all cases via
the Berry phase technique \cite{polariz,nota} using
a 16-point Monkhorst-Pack \cite{MP} k-point mesh in the $a$-plane
direction and 10 point uniform mesh in the $c$ direction.
For the supercell calculation, we have employed
superlattices  including typically 16-20 atoms,
oriented along (0001) for wurtzites, (111) or (100) 
for zincblende, (100)
for the  NaCl structure, and (100) for cubic 
perovskites. (Note that wurtzite has two independent components 
of the  dielectric tensor, and the one we  are actually calculating
is that  along the polar axis.)   In the supercell calculations,
downfolded  meshes were used comprising
 12 k-points for wurtzite and zincblende,
8 points for the NaCl-structure, and 12 points for perovskites.
 The  cation 
sublattice displacements $\delta$ are typically  0.05-0.1 \% of 
the bond length. Ionic relaxation is never allowed, so that the 
response is purely electronic.
All the calculations are performed at the theoretical 
lattice constants, that are reported in Table \ref{tab0}.

\begin{figure}[h]
\unitlength=1cm
\begin{center}
\begin{picture}(6,6.3)
\put(-0.6,-0.0){\epsfysize=6.3cm
\epsffile{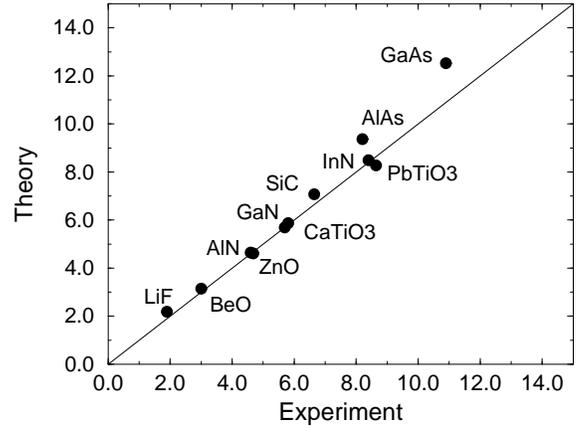}}
\end{picture}
\end{center}
\caption{Present theoretical 
$\einf$ vs.  experimental values.}
\label{fig1}
\end{figure}

\begin{figure}[h]
\unitlength=1cm
\begin{center}
\begin{picture}(6,6.3)
\put(-0.6,-0.0){\epsfysize=6.3cm
\epsffile{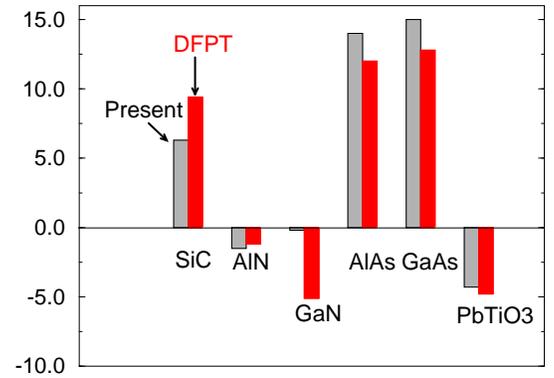}}
\end{picture}
\end{center}
\caption{Relative percental error in $\einf$ for DFPT and the present
method (see text).}
\label{fig2}
\end{figure}

With the above reported theoretical ingredients, we obtained the
 theoretical values of $\einf$  listed in Table
\ref{tab1} together with DFPT and experimental values, and  plotted in
Fig. \ref{fig1} versus the experimental values
for the different materials. The general level of
 agreement seems quite good
 on the scale of Fig. \ref{fig1}. To give a closer view,
in Fig. \ref{fig2} we display the relative percental deviation
of the theoretical  $\einf$ with respect to experiment, both for
our method and DFPT.

\paragraph*{Discussion -- }
The main content of Fig. \ref{fig2} is that
 DFPT and the present method agree quite closely
(the deviation for GaN is probably due to a different
 treatment of the Ga 3$d$ electrons).
It thus appears  that deviations from experiment are not
related to the specific method used, but are likely to be a token of
the underlying density functional formalism.  Most theoretical values 
are overestimates of the experimental data, the main  exception 
being the 4 \% underestimate for PbTiO$_3$, both in DFPT \cite{rabe}
(at the 
experimental  lattice constant)  and in the present method (at the
theoretical lattice constant). 
This is possibly due to the uncertainties  in the  experimental values,
 which are in fact plasmon-pole
extrapolations \cite{zhong}   to optical $\omega$=0 of values measured
in the visible at  relatively high temperature, while the calculation
is at zero  temperature and zero $\omega$ (in fact, this holds also 
for CaTiO$_3$).  Indeed  the situation for 
perovskites (even in the paraelectric phase) is far from settled in
general,  other recent linear response  results of another group
\cite{kra1,kra2} overestimate experiment considerably :
the  $\einf$ reported for cubic SrTiO$_3$ is 6.63  compared
with 4.69 experimental,\cite{kra1} and  
also 6.63 compared with 5.18 experimental for cubic KNbO$_3$.\cite{kra2}

It should be mentioned that $\einf$ in our scheme 
is actually a finite-$q$ value due to the finiteness of the 
simulation supercell, and this may cause some additional 
deviation as compared to DFPT. However, in our supercells the 
minimum $q$ is quite small ($\sim 0.03$ bohr$^{-1}$), and inspection
of the typical  structure \cite{walter}
 of $\einf$ as a function of $q$ reveals that
the deviations to be expected are in the order of 1 \%.

In closing, we discuss the case of 
 non-polar materials: in such a case,
  the procedure outlined above does not apply, 
since  no macroscopic  polarization can occur in a system 
containing only a non-polar material.
However, since the internal fields in the superlattice layers are
{\it proportional to $\Delta P$},  we can set up a superlattice 
by alternating layers of the unpolarized material of 
interest and layers of some appropriate polarized material. The 
latter effectively function as a polarization supply for the
unpolarized layer. In practice,  to compute the dielectric
constant of Si,  
we first calculate the dielectric constant and zero-field
polarization for some  polarized material, say SiC in the
wurtzite structure. We  then stack  along the (111) direction of
zincblende (i.e. (0001) of wurtzite),  a superlattice such as
[.../SiC/Si/SiC/Si/...].  Since SiC is polarized and we use 
periodic boundary conditions, the interface charge is
\be \label{eq.einf2}
  s_{\rm int} = \pm 2\, P_{\rm SiC}^{(0)}/
(\varepsilon^{\infty}_{\rm SiC} + 
\varepsilon^{\infty}_{\rm Si}),
\ee
whence the dielectric constant of Si is trivially extracted.
SiC should have 
the in-plane lattice constant of Si in order  to avoid strain effects
in the Si layer. Clearly its calculated properties in this
specific, hypotetical  realization are irrelevant: what counts is that it
provides the polarization to create an interface charge and  a
depolarizing field inside Si.  The above scheme, it turns out,
 is more of conceptual interest than of practical use for non-polar 
solids: interface states occur fairly easily  at heterovalent junctions,
spoiling the applicability of Eq. (\ref{eq.einf2}).  Applying the scheme  
 in practice to  non-polar solids will require  quite some trial and
error to identify a ``clean'' interface, and we did not pursue 
this further here.

\paragraph*{Summary and acknowledgements -- }
In conclusion, our results indicate that the polarization-based method
can produce theoretical  dielectric constants within $\sim$5 to 10 
\% of experiment, and is  as accurate as DFPT. While much more limited 
in its general scope, our method appears to be a useful alternative to 
DFPT for this kind of calculations. 

We thank Elena Manca for help in the perovskite calculations, and 
acknowledge special support from INFM within the PAISS program.


\narrowtext

\begin{table}[ht]
\caption{Theoretical structural parameters for the materials being studied.} 
\begin{tabular}{lcccc}
   & Structure    & $a_0$ (bohr)&$c_0/a_0$ &$u_0$   \\ \hline 
GaAs & zincblende & 10.60&         &        \\ 
AlAs & zincblende & 10.62&         &         \\ 
LiF & rocksalt& 7.50  &         &          \\
AlN & wurtzite & 5.82 & 1.619  & 0.380          \\
GaN & wurtzite & 6.04 & 1.634  & 0.376           \\
InN & wurtzite & 6.66  & 1.627  & 0.377           \\
BeO & wurtzite & 5.00 & 1.610    & 0.377           \\
ZnO & wurtzite & 5.98 & 1.616  & 0.376         \\
PbTiO$_3$ & cubic & 7.34 & & \\
CaTiO$_3$ & cubic & 7.20 & &
\end{tabular}
\label{tab0}
\end{table}

\vbox{\begin{table}[h]
\caption{Electronic dielectric constants of several polar 
insulators. Our calculated values are compared with theoretical
 DFPT values and with experiment.}
\begin{tabular}{lccc}
 &   Present & DFPT & Exp. \\
\hline
GaAs &        12.53 & 12.3$^a$  &   10.9$^f$   \\
AlAs &        9.37  & 9.2$^a$   &   8.2$^f$ \\
InN   &       8.49  &       &   8.40$^f$   \\ 
w-SiC  &      7.07  & 7.28$^b$  &   6.65$^b$  \\
GaN    &      5.69  & 5.41$^c$  &   5.70$^g$\\
ZnO    &      4.65  &       &   4.60$^f$ \\
AlN    &      4.61  & 4.62$^d$  &   4.68$^h$\\
BeO    &      3.15  &       &   2.99$^i$ \\
LiF    &      2.19  &       &   1.96$^f$\\
PbTiO$_3$ &   8.28  & 8.24$^e$ & 8.64$^j$ \\
CaTiO$_3$ &   5.87  &       &    5.81$^j$
\end{tabular}
$^a$ Ref. \onlinecite{dfpt2}, 
$^b$Ref.~\onlinecite{karch}, 
$^c$Ref.~\onlinecite{karch2}, 
$^d$Ref.~\onlinecite{pasq},
$^e$Ref.~\onlinecite{rabe},
$^f$Ref.~\onlinecite{expeps},
$^g$Ref.~\onlinecite{ref.f}, 
$^h$Ref.~\onlinecite{ref.g}, 
$^i$Ref.~\onlinecite{ref.h}, 
$^j$Ref.~\onlinecite{zhong}. 
\label{tab1}
\end{table}}

\end{multicols}

\begin{thebibliography}{999}
\bibitem{nota2}
For layers of different thicknesses, Eq. \protect\ref{eq.einf} 
 reads 
\be
\mid\!s_{\rm int}\!\mid =  \Delta P^{(0)}\,
( l_1 + l_2 )/ (l_2 \einf_1 + l_1 \einf_2),
\nonumber
\ee
and is obtained accounting for the general  periodic boundary
conditions $l_A E_A + l_B E_B = 0$
in the determination of $\Delta E$ and hence of  $s_{\rm int}$.
\end{thebibliography}
\end{document}